\documentclass[jgrga]{agutex}

  \usepackage[dvips]{graphicx}
\usepackage{amsmath}
\usepackage{amssymb}
\usepackage{amsfonts}
\usepackage{bbm}

\authorrunninghead{SCHUMANN ET AL.}

\titlerunninghead{HEAVY-TAILED PROCESSES WITH LONG-RANGE MEMORY}

\authoraddr{A.~Y. Schumann,
Complexity Science Group, Department of Physics and Astronomy, University of
Calgary, 2500 University Dr. NW, Calgary, Alberta T2N 1N4, Canada.
(ay.schumann@ucalgary.ca)}

\authoraddr{J. Davidsen,
Complexity Science Group, Department of Physics and Astronomy, University of
Calgary, 2500 University Dr. NW, Calgary, Alberta T2N 1N4, Canada.
(davidsen@phas.ucalgary.ca)}

\begin{document}

%% ------------------------------------------------------------------------ %%
%
%  TITLE
%
%% ------------------------------------------------------------------------ %%

\title{Extreme value and record statistics in heavy-tailed processes with long-range memory}
%
% e.g., \title{Terrestrial ring current:
% Origin, formation, and decay $\alpha\beta\Gamma\Delta$}
% You may use \\ to break the title over several lines.

%% ------------------------------------------------------------------------ %%
%
%  AUTHORS AND AFFILIATIONS
%
%% ------------------------------------------------------------------------ %%

%Use \author{\altaffilmark{}} and \altaffiltext{}

% \altaffilmark will produce footnote;
% matching altaffiltext will appear at bottom of page.
% May use \\ to start a new line.

\authors{Aicko Y. Schumann\altaffilmark{1},
Nicholas R. Moloney\altaffilmark{1,2} and J\"orn Davidsen\altaffilmark{1}}

\altaffiltext{1}{Complexity Science Group, Department of Physics and
Astronomy, University of Calgary, 2500 University Dr. NW, Calgary, Alberta T2N
4L1, Canada.}

\altaffiltext{2}{Max Planck Institute for the Physics of Complex Systems, N\"othnitzer
Str. 38, 01187 Dresden, Germany.}

%% ------------------------------------------------------------------------ %%
%
%  ABSTRACT
%
%% ------------------------------------------------------------------------ %%

% >> Do NOT include any \begin...\end commands within
% >> the body of the abstract.

\begin{abstract}
Extreme events are an important theme in various areas of science because of
their typically devastating effects on society and their scientific
complexities. The latter is particularly true if the underlying dynamics does
not lead to independent extreme events as often observed in natural systems.
Here, we focus on this case and consider stationary stochastic processes that
are characterized by long-range memory and heavy-tailed distributions, often
called fractional L\'evy noise. While the size distribution of extreme events
is not affected by the long-range memory in the asymptotic limit and remains a
Fr\'echet distribution, there are strong finite-size effects if the memory
leads to persistence in the underlying dynamics. Moreover, we show that this
persistence is also present in the extreme events, which allows one to make a
time-dependent hazard assessment of future extreme events based on events
observed in the past. This has direct applications in the field of space
weather as we discuss specifically for the case of the solar power influx into
the magnetosphere. Finally, we show how the statistics of records, or
record-breaking extreme events, is affected by the presence of long-range
memory.
\end{abstract}

%% ------------------------------------------------------------------------ %%
%
%  BEGIN ARTICLE
%
%% ------------------------------------------------------------------------ %%

% The body of the article must start with a \begin{article} command
%
% \end{article} must follow the references section, before the figures
%  and tables.

\begin{article}

%% ------------------------------------------------------------------------ %%
%
%  TEXT
%
%% ------------------------------------------------------------------------ %%

\section{Introduction}
\label{sect:introduction}

History often turns on extreme events --- rare occurrences of extraordinary
nature --- be they man-made or natural. Examples are global financial crises,
military strikes, radical political events, or natural disasters such as
floods, droughts and earthquakes. Extreme events are often associated with
catastrophes, and the word `extreme' is sometimes substituted by `freak' to
suggest something unnatural and undesirable. Generally, the economic and
social consequences of extreme events are a matter of enormous concern. In
particular, the ever increasing economic and human losses from natural hazards
underscore the urgency for improved understanding of extreme events to develop
effective strategies to reduce their impact. 

The surprisingly high likelihood of extreme events is actually a key attribute
of many complex systems, in both natural and man-made environments (see, e.g.,
\citep{albeverio,bunde,embrechts,galambos_ed,sornette}).
In particular, we know that records must be broken in the future, so if a
flood design is based on the worst case of the past, then we are still not
prepared for all possible floods in the future.  The classic approach to
studying the probability of extreme events has been to assume independent and
identically distributed (iid) event sizes. This has led to a powerful
statistical theory (see, e.g.,
\citep{embrechts,Coles_Extreme_Book,ExtremeValueTheory_deHaan}), which has been
successfully applied in many cases (see, e.g.,
\citep{easterling00,glaser04,brink08}).
The latter is also related to the fact that some
parts of the theory --- for example, the limit distributions of block maxima
--- can be extended to a wide class of dependent stationary stochastic
processes and their associated time
series~\citep{berman64,leadbetter,leadbetter88,samorodnitsky04}. However, we
are still far away from a general understanding of extreme events generated by
such dependent processes, which are abundant in nature. Complicating factors
typically include slow convergence and strong finite-size
effects~\citep{GyorgyiETAL:2008}, as well as nonlinear
correlations~\citep{BogachevEichnerBunde:2007}.

Strikingly, one often encounters dependence in the form of long-range
persistence in recordings taken from natural systems. Persistence is defined
as the tendency that subsequent values in a time series are similar: large
values tend to be followed by large values and small values tend to be
followed by small values. Such behavior has been reported for water levels in
rivers~\citep{hurst51} and in river runoff
records~\citep{Kantelhardt_RiverRunoff:2003,Kantelhardt:2006}, in
climatological temperature
recordings~\citep{Koscielny-Bunde:1998,pelletier97,huybers06} and for
temperature fluctuations in oceans~\citep{monetti02}, as well as in marine
data~\citep{Roman:2008}, to name only a few examples. 

The theoretical studies of the extreme value statistics of stationary
stochastic processes exhibiting long-range persistence have mainly
focused on processes with Gaussian distributed event sizes. In this
and related cases, the limit distribution of the block maxima is the
same as in the iid case --- a Gumbel
distribution~\citep{berman64,eichner06}. Similar results have been
obtained for beta-distributed random variables, in which case the
Gumbel distribution is replaced by the Weibull
distribution~\citep{moloney08}. Meanwhile, for distributions with
power-law tails the block maxima for iid events are Fr\'echet
distributed~\citep{embrechts}, and this is even true for a certain
class of dependent sequences~\citep{leadbetter,leadbetter88}. 
Power-law tails imply that there is no ``typical'' event size and that
indeed event sizes vary over many orders of magnitude. Many
natural systems even obey distributions of Pareto-like (power-law-like)
\emph{heavy} tails, such that the second moment of the distribution is not
defined. In space
physics, such heavy-tailed distributions have been used to model
e.g. magnetic field line transport~\citep{PommoisZimbardoVeltri:1998},
fluctuations in various solar wind
parameters~\citep{HnatChapmanRowlands:2003,BrunoETAL:2004,zaslavsky08}
and auroral indices~\citep{watkins05,zaslavsky08}. Other examples
include, size distributions in geology~\citep{HeavyTailGeology_Caers:1999b}, meteorology
\citep{taqqu87}, or sediments \citep{painter94}.

Here, we study the extreme value statistics of exactly those stationary
stochastic processes that are characterized by persistent (or anti-persistent)
long-range memory and heavy tails. These processes are sometimes called
fractional L\'evy noise or linear stable fractional noise ---
see~\citep{Watkins:2009,CorrelatedLevy_SolarWinds_Moloney2010} for discussions and further references.
Specifically, we consider those symmetric $\alpha$-stable distributed
processes ($0<\alpha<2$) that are characterized by a self-similarity index $H$ with
$0<H<1$. These are particularly relevant in the context of the solar wind as
discussed, for example, by~\cite{CorrelatedLevy_SolarWinds_Moloney2010} for the energy influx into the
magnetosphere, which is captured by the Akasofu $\epsilon$ parameter. While it
is known that the limit distribution of the block maxima remains a Fr\'echet
distribution for all $\alpha$ and $H$~\citep{samorodnitsky04}, we
systematically quantify the finite size corrections. As expected, they are
particularly pronounced for strong persistence but they also play a role in
the presence of anti-persistence. Moreover, we show that the conditional block
maxima distributions deviate significantly from the unconditional distribution
for small block sizes. This history dependence allows one to make a
time-dependent hazard assessment of the value of the next block maximum based
on the previous one. We also show how the statistics of records, or
record-breaking extreme events, is affected by the presence of long-range
memory and, thus, extend very recent results for the Gaussian distributed
case~\citep{RecordsCorrelatedGaussianTemperatures_Newman:2010}.  Finally, we
apply the time-dependent hazard assessment and the record analysis to the
$\epsilon$ time series derived from ACE spacecraft measurements for the years
$2000-2007$. 

The paper is organized as follows: First we review basic properties and limit
theorems of (maximum) extreme value statistics in
Sect.~\ref{sect:extreme_value_statistics}, followed by record statistics in
Sect.~\ref{sect:record_statistics}. We then describe $\alpha$-stable- or
L\'evy distributions and discuss the generation of stationary symmetric
$\alpha$-stable distributed processes with long-range memory in
Sect.~\ref{sect:alpha_stable}. The presentation of our numerical results can
be subdivided into two parts. While we focus on estimating the extreme value
statistics of these processes including simple block maxima and conditional
block maxima in Sects.~\ref{sect:results_block_maxima} and
\ref{sect:results_cond_block_maxima}, their record statistics is presented in
Sect.~\ref{sect:results_records}. Section~\ref{sect:solar_application}
discusses the application of the introduced methodology to the solar power
influx into the Earth's magnetosphere. Finally we conclude in
Sect.~\ref{sect:conclusion}.

\section{Classical Extreme Value Statistics}
\label{sect:extreme_value_statistics}

In classical extreme value statistics one considers sets of
independent identically distributed random variables
$\{X_1,X_2,\ldots,X_n\}$, each drawn from the same cumulative
distribution function $F(x)$. $F(x)$ can typically be represented by a
unique probability density function $P(x)$.  A time series
$\{x_i\}_{i=1,\ldots,n}$ can then be understood as one possible
realization of the set of random variables,
$\{X_1=x_1,X_2=x_2,\ldots,X_n=x_n\}$, following this density function
$P(x)$. The distribution of the (block) maximum or extreme value
$M_n=\max\{X_1,\ldots,X_n\}$
\footnote{Since $\min\{X_1,\ldots,X_n\}=\max\{-X_1,\ldots,-X_n\}$, the minimum can be mathematically treated in the same way.}
is given by

\begin{equation}
   \begin{aligned}
      Pr(M_n\le m) &= Pr(X_1\le m, \ldots, X_n\le m)\\
                   &= \prod_{i=1}^n Pr(X_i\le m)=F^n(m)\ .
   \end{aligned}
   \label{eq:maximum_distribution}
\end{equation}
The \emph{Fischer-Tipett-Gnedenko theorem} 
states that if there exists a renormalization sequence $\{a_n, b_n\},
a_n\in\mathbbm{R}^+$ and $b_n\in\mathbbm{R}$, such that

\begin{equation}
   \lim_{n\to\infty} Pr\left(\frac{M_n-b_n}{a_n}\le m\right) \stackrel{d.}{=}
   G(m)\ ,
   \label{eq:Fischer_Tipett_Gnedenko}
\end{equation}

where $G(m)$ is non-degenerate and $d.$ means convergence in distribution, then the
asymptotic limit distribution $G(m)$ belongs to one of the
following three function families~\citep{Coles_Extreme_Book}:

\begin{subequations}
\begin{align}
   G_\text{Gumbel}(m) &= \exp\left\{-\exp\left\{-\left(\frac{m-b}{a}\right)\right\}\right\}
   \label{eq:Gumbel}\\
   G_\text{Fr\'echet}(m) &= \begin{cases}
                              0 &\ \ \ : m\le b\\
                              \exp\left\{-\left(\frac{m-b}{a}\right)^{-\alpha}\right\}
			      &\ \ \ : m>b
			    \end{cases}
   \label{eq:Frechet}\\
   G_\text{Weibull}(m) &= \begin{cases}
			       \exp\left\{-\left(-\left(\frac{m-b}{a}\right)\right)^\alpha\right\} &: m<b\\
			       1 &: m\ge b.
			    \end{cases}
			    \label{eq:Weibull}
\end{align}
   \label{eq:ev_dists}
\end{subequations}

Here, $a\in\mathbbm{R}^+,b\in\mathbbm{R}$ and $\alpha>0$. As evident
from Eq.~\eqref{eq:maximum_distribution}, the exact limiting
distribution is determined by $F(x)$ and, thus, $P(x)$.

If, for instance, $P(x)$ corresponds to the normal distribution
or the exponential distribution, the
limit distribution 
of the (block) maxima or extreme values 
follows
a \emph{Gumbel} distribution (Eq.~\eqref{eq:Gumbel}).
In cases where $P(x)$ exhibits power law right tails (Pareto-like
tails), i.~e., $\overline{P}(x)\sim L(x)x^{-1-\alpha}$ where
$\overline{P}(x)$ is the tail of the distribution $P(x)$, $L(x)$ is
some slowly varying function, the limit distribution of the (block)
maximum or extreme value approaches the \emph{Fr\'echet} distribution
(Eq.~\eqref{eq:Frechet}) with the same $\alpha>0$.
Finally, the limit distribution of the
maximum falls into the \emph{Weibull}
class for some distributions with bounded right tails, such as
the beta distribution (Eq.~\eqref{eq:Weibull}).
Note that the convergence to one of the three distributions in
Eq.~\eqref{eq:ev_dists} requires the condition in
Eq.~\eqref{eq:Fischer_Tipett_Gnedenko} to be valid.
For example, the condition in Eq.~\eqref{eq:Fischer_Tipett_Gnedenko} is not 
generally satisfied for the Poisson distribution, the geometric distribution, 
or the negative binomial
distribution~\citep{ExtremeValueTheory_deHaan,Modelling_Extremal_Events:2003}.

All three extreme value distributions in Eqs.~\eqref{eq:ev_dists} can be
written in a compact form in terms of the generalized extreme value
distribution (GEV)

\begin{equation}
   G(z) = \exp\left\{ -\left[ 1+\xi\left( \frac{z-\mu}{\sigma} \right)
   \right]_+^{-1/\xi} \right\}
   \label{eq:gev}
\end{equation}

where $+$ indicates the constraint $\left[ 1+\xi\left( \frac{z-\mu}{\sigma}
\right) \right]>0$. The parameters are the \emph{shape parameter} $\xi$, $
\xi\in\mathbbm{R}$, the \emph{scale parameter}
$\sigma$, $\sigma\in\mathbbm{R}^+$, and the \emph{location parameter}
$\mu$, $\mu\in\mathbbm{R}$. The shape parameter $\xi$ determines the distribution
family: $\xi=0$ (interpreted as $\lim_{\xi\to 0}$) indicates the Gumbel
class (Eq.~\eqref{eq:Gumbel}), $\xi>0$ the Fr\'echet class
(Eq.~\eqref{eq:Frechet}), and $\xi<0$ the Weibull class
(Eq.~\eqref{eq:Weibull}).

In this paper, we exclusively focus on symmetric $\alpha$-stable
distributed processes with long-range memory, for which the block
maxima asymptotically converge in distribution to the Fr\'echet
extreme value distribution in
Eq.~\eqref{eq:Frechet}~\citep{samorodnitsky04}.

\section{Classical Record Statistics}
\label{sect:record_statistics}

The field of record statistics is closely related to extreme value statistics
since records can be considered as a special type of extreme values: Records
are simply record-breaking extreme values, i.e., the sequence of extreme
values for increasing $n$. Studies of record values and record times started
with the pioneering work by \cite{RecordsFirst_Chandler:1952} followed by
important contributions by \cite{Records_Reny:1962}. For a comprehensive
overview and a historic summary we refer to \citep{RecordBreaking_Glick:1978}
and \citep{RecordsTHEBOOK_Nevzorov:2001} and references therein. The classical
theory of records is based on the assumption of iid random variables and has
been successfully extended and applied to many natural systems, including
earthquakes~\citep{RecordsEarthquake_Davidsen:2006,RecordsNetworks_Davidsen:2008,RecordsEarthquake_VanAalsburg:2010,vasudevan10,peixoto10},
climate
dynamics~\citep{schmittmann99,benestad03,RecordsGlobalWarming_Redner:2006,benestad08,RecordsCorrelatedGaussianTemperatures_Newman:2010,wergen10},
hydrology~\citep{vogel01} and evolution~\citep{sibani09}.

\begin{figure}[tb]
   \centering
   \includegraphics[width=0.8\hsize]{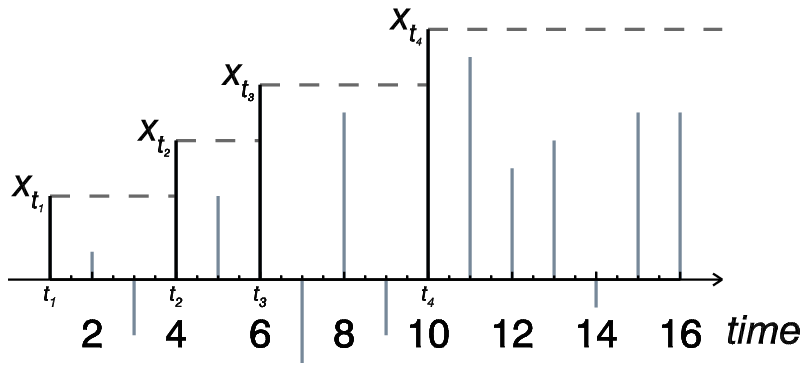}
   \caption{Illustration on the definition of records in a realization
   $\{x_i\}_{i=1,\ldots,R}$ of the process $\{X_i\}_{i=1,\ldots,R}$.  Four
   records (black), $r_1=x_{t_1=1}$, $r_2=x_{t_2=4}$, $r_3=x_{t_3=6}$,
   $r_4=x_{t_4=10}$, are identified in the shown fragment.
   %The waiting times
   %between consecutive records, $t_{k+1}-t_k$, are marked by blue dashed
   %lines. \JD{are waiting times relevant?}
   }
   \label{fig:records_illustration}
\end{figure}

Specifically, let again $\{X_1, X_2, \ldots, X_R\}$ be a sequence of
iid~random variables with a common continuous distribution $F(x)$. One
possible realization of the set of random variables may then be given by the
time series $\{x_i\}_{i=1,\ldots,R}=\{x_1,x_2,\ldots,x_R\}$.  We denote $X_m$
as an (upper) record if $X_m=\max\{X_1,\ldots,X_m\}$ for $m \leq R$
\footnote{Similarly, one can define a lower record $X_m=\min\{X_1,\ldots,X_m\}$.}. 
As illustrated in Fig.~\ref{fig:records_illustration}, the set of records for
a given realization is trivially ordered and the index $m$ at which the $k$-th
record occurs is defined as the $k$-th record time, $T_k$.  Obviously, $T_1=1$
and $T_{k\ge 2}=\min\{j\le R:X_j>X_{T_{k-1}}\}$. Thus, $T_{k>1}$ as well as
the number of records after a time $t\leq R$ in a block of size $R$, $N_t$,
are random variables themselves. 
%We further denote by $R_k$ the random
%variable of the $k$-th record ($1\le k\le N_R$).

Interestingly, many statistical properties of records are identical for
all sequences of iid~random variables $\{X_i\}_{i=1,\ldots,R}$ and, thus, independent 
of the distribution $F(x)$. This includes the expected number of records and 
their variance. To see this, it is important to realize that each $X_{k\leq m}$ 
has the same probability of being the maximum, which is simply given by 

\begin{equation}
   P(X_k=\max\{X_1,\ldots,X_m\})=1/m\ .
   \label{eq:record_rate}
\end{equation}

Thus, this is in particular the probability that $X_m$ is a record, which 
allows one immediately to calculate the expected number of records after a
time $t$ and their variance~\citep{RecordBreaking_Glick:1978}:
\begin{subequations}
   \begin{align}
      E(N_t) &=\sum_{m=1}^t P(X_m=\max\{X_1,\ldots,X_m\})\notag\\
      &= \sum_{m=1}^t 1/m =
      \ln(t)+\gamma+\mathcal{O}(1/t)\label{eq:expected_number_records}\\
      \text{Var}(N_t)&=\sum_{m=1}^t \frac{1}{m}-\sum_{m=1}^t
      \frac{1}{m^2}\notag\\
      &=\ln(t)+\gamma+\mathcal{O}(1/t)-\pi ^2/6\ ,
   \end{align}
\end{subequations}
where $\gamma = 0.577\ldots$ is the Euler-Mascheroni constant.

Note, that there are other properties of records, as for instance the
distribution of occurrence times of the $k$-th
records~\citep{RecordBreaking_Glick:1978}, that are also independent of the
distribution of $\{X_i\}_{i=1,\ldots,R}$ but will not be further discussed
here. In contrast to the above distribution-independent properties, results
involving actual record values do generally dependent on the distribution
$F(x)$ \citep{RecordsTHEBOOK_Nevzorov:2001,RecordsGlobalWarming_Redner:2006}. 

\section{Heavy-tailed processes with long-range memory}
\label{sect:alpha_stable}

A prototype of stationary stochastic processes with heavy-tailed distributions
are $\alpha$-stable processes. A random variable $X$ is called stable if any
set of independent copies of $X$, $\{X_1,\ldots,X_n\}$, obeys

\begin{equation}
   X_1+\ldots+X_n \stackrel{d.}{=}c_nX+d_n
   \label{eq:stable}
\end{equation}
for some $c_n\in\mathbb{R}^+$ and $d\in\mathbbm{R}$
\citep{ExtremeFinanceInsurance_Embrechts:1994}. This property implies that the shape 
of the distribution remains invariant if sums over the random variable are 
considered. If $d_n=0$, $X$ is referred to as
strictly stable, and $X$ is symmetric if $X\stackrel{d.}{=}-X$. 
A generalized central limit theorem proves
that all distributions that have a probability density function $P(x)$ with
an associated Fourier transform or characteristic function of the form
% HOW DO YOU GET A WIDE EQUATION IN AGU STYLE ?????
\begin{equation}
   \varphi_{\alpha,\beta,\delta,\gamma}(t) = \begin{cases}
      \exp\left\{it\delta
      -\gamma^\alpha|t|^\alpha\left(1-i\beta\;\text{sgn}(t)\tan\left(\dfrac{\pi\alpha}{2}\right)\right)\right\} &: \alpha\ne 1\\
      \exp\left\{it\delta -\gamma|t| \left(1+\dfrac{2}{\pi}i\beta\;\text{sgn}(t)\ln
      |t|\right)\right\}
      &:
      \alpha =1
   \end{cases}
   \label{eq:alpha_stable_characteristic_function}
\end{equation}
are stable \citep{NonGaussianRandomProcesses:1994}. In particular, the 
constants $c_n$ are not arbitrary but scale as $n^{1/\alpha}$ with $\alpha\in (0,2]$, and hence, 
the corresponding distributions are called $\alpha$-stable ($\alpha$S) distributions.
In Eq.~\eqref{eq:alpha_stable_characteristic_function}, the 
\emph{characteristic exponent} or
\emph{index of stability} $\alpha\in (0,2]$, the \emph{skewness parameter}
$\beta\in[-1,1]$, and the \emph{scale parameter} $\gamma\in\mathbbm{R}^+$
describe the shape of the distribution while $\delta\in\mathbbm{R}$ is the
\emph{location parameter}. Since many of these $\alpha$S distributions
do not posses a finite mean and standard deviation, these quantities are typically not
used to characterize these distributions.

For symmetric $\alpha$-stable (S$\alpha$S) distributions, the skewness parameter
obeys $\beta=0$. In particular, the Gaussian distribution $\mathcal{N}(\mu,\sigma^2)$ is S$\alpha$S
with $(\alpha=2,\beta=0,\gamma=\sigma/\sqrt{2},\delta=\mu)$. Throughout this paper we
will only focus on S$\alpha$S distributed time series and set without loss of
generality the location parameter $\delta$ to zero. We will further
normalize our time series to ensure $\gamma=1$ (see
Sect.~\ref{sect:numerics}). Hence,
Eq.~\eqref{eq:alpha_stable_characteristic_function} reduces to the simple form
\begin{equation}
   \varphi_{\alpha}(t) = \exp\left\{-|t|^\alpha\right\}\ .
   \label{eq:centered_symmetric_alpha_stable_characteristic_function}
\end{equation}
While even in this case an analytical form of $P(x)$ is typically not known,
the asymptotic behavior for $\alpha\in (0,2)$ is characterized by
\emph{Pareto-like} power-law tails with $\overline{P}(x)\sim x^{-1-\alpha}$
\citep{sornette}. 
Consequently, the variance of all
non-Gaussian $\alpha$S distributions with $\alpha<2$ is undefined. In
addition, the distribution's mean value is undefined for $\alpha<1$. 

\subsection{Long-range memory in symmetric $\alpha$-stable processes}
\label{sect:corr_alpha_stable}

For a sequence of identically but \emph{not} independently distributed random
variables $\{X_1, X_2, \ldots, X_n\}$ with a common continuous distribution
$F(x)$, the associated realizations or stationary time series
$\{x_i\}_{i=1,\ldots,n}=\{x_1,x_2,\ldots,x_n\}$ are typically characterized by
non-trivial correlations. One particular example is the presence of long-range
linear auto-correlations. These can be described by three equivalent
definitions, namely a power-law behavior in (i) the autocorrelation function
with correlation exponent $0<\gamma<1$, $C(s)\sim s^{-\gamma}$, (ii) the power
spectrum with spectral exponent $0<\beta<1$, $P(f)\sim f^{-\beta}$, as well as
(iii) the fluctuation function with (monofractal) fluctuation exponent $h(2)$, $F(s)\sim
s^{h(2)}$ for scales $s$, frequencies $f$, and $1-\beta=\gamma=2-2h(2)$; for a
comprehensive definition and discussion, see \citep{Schumann_book:2011} and
references therein.

While such long-range linear auto-correlations imply long-range persistence,
the reverse is not necessarily true. In particular, the above definitions
require that a finite variance exists which is not the case for
$\alpha$-stable processes with $\alpha < 2$ as discussed above. Thus, one has to use an
alternative way to quantify persistence and long-range memory in general for
such processes. One possibility is to define persistence based on the closely
related phenomenon of self-similarity. A (continuous) stochastic process
$X=\{X(t), t\in\mathbbm{R}\}$ is called self-similar with a self-similarity
parameter $H>0$ if $\forall \kappa>0$, $t\in\mathbbm{R}$,
\citep{SelfSimilarProcesses_Embrechts:2002},
\begin{equation}
   \{X(\kappa t)\}\stackrel{d.}{=}\{\kappa^H X(t)\}\ .
   \label{eq:self_similarity}
\end{equation}
It can be further shown that for self-similar S$\alpha$S distributed processes
$X_{H,\alpha}$ with $0<H<1$, long-range persistence is achieved for
$H>1/\alpha$, while $H=1/\alpha$ corresponds to the uncorrelated case and
anti-persistence emerges for $H<1/\alpha$ \citep{Stoev:2004}. Thus,
persistent behavior can only be observed if $1 < \alpha < 2$ and we focus on
this range of $\alpha$ values in the following.

\subsection{Numerical simulation of self-similar symmetric $\alpha$-stable processes}
\label{sect:numerics}

There are two distinct algorithms typically used to generate stationary
self-similar S$\alpha$S-distributed  processes characterized by the exponents
$\alpha$ and $H$. One approach utilizes fractional calculus and derives the
fractional integral or fractional derivative $X_{\nu,\alpha}(t)$ of an
uncorrelated S$\alpha$S-process $X_\alpha(t)$ in Fourier space,
$\hat{X}_{\alpha}(\omega)=(2\pi)^{-1}\int_\mathbbm{R}X_\alpha(t^\prime)\exp\{-it^\prime\omega\}dt^\prime$,
\citep{Chechkin_Gonchar:2000},
\begin{equation}
   \hat{X}_{\nu,\alpha}(\omega) = \frac{\hat{X}_\alpha(\omega)}{(i\omega)^\nu},\ \nu=H-1/\alpha\ .
\end{equation}
Hence, this approach is very similar to the established Fourier-filtering
technique often used for generating correlated Gaussian noises, see e.~g.
\citep{Schumann_MF-CMA:2011}. Note that \cite{Chechkin_Gonchar:2000} restrict
their algorithm to $1\le\alpha\le 2$ \footnote{This limitation mainly comes
from the loss of validity of the Minkowski inequality for $\alpha<1$, which is
used in \citep{Chechkin_Gonchar:2000} to estimate the upper boundary of $H$.
It should nevertheless be  possible to generalize the algorithm to arbitrary
$\alpha\in(0,2]$. Moreover, one possible pitfall of the algorithm is that the
probability density of the generated self-similar S$\alpha$S process is not
explicitly renormalized to be the same as for the uncorrelated noise it is
initialized with. This has to be done manually.}. They further discuss why
high-frequency inaccuracies of fractional integration disturb numerical
simulations in the anti-persistent case, see also
\citep{FractionalIntegrationFFT_Chechkin:2001} for their comprehensive study
of fractional Gaussian integration.

The other algorithm is based on discretizing and numerically solving a
stochastic integral that defines a \emph{cumulative} self-similar S$\alpha$S process
$Y_{H,\alpha}(t)$, often called fractional L\'evy motion, 
\begin{equation}
   Y_{H,\alpha}(t) =
   C_{H,\alpha}^{-1}\int_\mathbbm{R}\left(\left(t-s\right)_+^{H-1/\alpha}-(-s)_+^{H-1/\alpha}\right)dY_\alpha(s)
   \label{eq:stoev_stochastic_integral}
\end{equation}
where $Y_\alpha(s)$ is (standard) S$\alpha$S motion and $C_{H,\alpha}$ is some norming
constant. The increments $X_{H,\alpha}(k)=Y_{H,\alpha}(k)-Y_{H,\alpha}(k-1)$
in the discretized version of Eq.~\eqref{eq:stoev_stochastic_integral} then
define a self-similar S$\alpha$S process $X_{H,\alpha}$ that we study here.
Note that the norming constant is chosen such that (i) the S$\alpha$S scaling
parameter is $\gamma=1$ as previously discussed for deriving
Eq.~\eqref{eq:centered_symmetric_alpha_stable_characteristic_function} from
Eq.~\eqref{eq:alpha_stable_characteristic_function} and (ii) the probability
density functions are equal for the independent and self-similar S$\alpha$S
process. A stochastic integral such as in
Eq.~\eqref{eq:stoev_stochastic_integral} can be discretized and efficiently be
solved by exploiting the convolution theorem and employing a fast Fourier
transform, for further details we refer to \citep{Stoev:2004}
\footnote{A similar supposedly faster algorithm was suggested later
\citep{FasterStoev_Wu:2004}, but not tested by us.}.

In the following, we present results for self-similar S$\alpha$S processes
using the latter algorithm~\citep{Stoev:2004} but we obtain very similar
results using the former one~\citep{Chechkin_Gonchar:2000} over the range of
its established validity. These processes are parametrized by the two
parameters $H$ and $\alpha$ ($\gamma=1,\delta=\beta=0$ in
Eqs.~\eqref{eq:alpha_stable_characteristic_function} and
~\eqref{eq:centered_symmetric_alpha_stable_characteristic_function}). Two
further parameters $m$ and $M$ are required by the used algorithm. They
control the mesh-size (intermediate points) and the kernel of the fast Fourier
transform, respectively. We chose the parameters $m=64$ and $M=48576$ and
generate time series of length $n=1000000$. This choice of $m$ and $M$ with
$mM \ll n$ is motivated by our observation that an undesired crossover in the
scaling behavior of different parameters
appears at scales of the order $mM$
\footnote{This crossover is not present in the algorithm by \cite{Chechkin_Gonchar:2000}}.
To the best of our knowledge, this has not been noticed before. Indeed, much
shorter time series with $n+M = 2^{14}$, $m_\text{max}=256$, and
$M_\text{max}=6000$ were studied by \citep{Stoev:2004} who concluded that
larger values of $m$ should be chosen for $0<\alpha\le 1$ while smaller $m$
are preferable for $1<\alpha\le 2$ and small $M$. This is clearly not
supported by our simulations for much larger values of $n$
\footnote{We also noted a number of artifacts in the generated time series for
values $\alpha<1$ which we do not consider in this paper. One might be able to
resolve these artifacts by significantly increasing the mesh size parameter
$m$. However, doubling $m$ results in practically doubling the memory
requirements unless the series lenght $n$ is reduced accordingly. We plan on
testing this more systematically in the future.}, 
which are necessary to estimate extreme value and record statistics of
self-similar S$\alpha$S processes reliably
\footnote{Note further that the values were chosen to ensure that $m(M+n)$ is
an integer power of $2$.}.

For each parameter set $(\alpha,H)$ we generate $N_\text{conf}=1500$
long-range persistent data sets
\footnote{Note that significant computational resources are required.  For a
single parameter set $(\alpha,H)$, about $15$GB are necessary to store the
`double' raw data. Since the computation of the involved Fourier transform
requires to allocate arrays of size $m(M+n)$ and there is a noticeable
overhead for computation, main memory requirements become an issue very
quickly as $m$ and $M$ grow.}. 
For surrogate testing we destroy correlations by shuffling those data sets
using a Mersenne-Twister-19937 pseudo-random number generator
\citep{MersenneTwister} which was also used to generate white Gaussian noise
being the basis for the uncorrelated S$\alpha$S noise \footnote{Uncorrelated
S$\alpha$S distributed random numbers can be generated from uncorrelated
Gaussian distributed random numbers following
\citep{StableRandomNumbers_Chambers:1976,StableRandomNumbers_Chambers:1987}.}.

\section{Effects of long-range memory on the statistical properties of symmetric $\alpha$-stable processes}
\label{sect:results}

\subsection{Extreme value statistics }
\label{sect:results_block_maxima}

\begin{figure}[tb]
   \centering
      \includegraphics[width=\hsize]{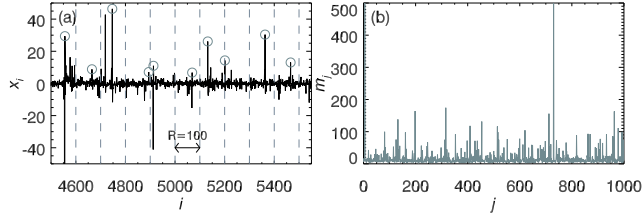}
      \caption{Definition of a sequence of block maxima 
      $m_j=\max\{x_{jR},\ldots,x_{j(R+1)-1}\}_{j=0,1,\ldots,[N/R]-1}$
      [partially shown in (b)] by considering disjunct blocks of size $R=100$ in the
      underlying time series $\{x_i\}_{i=0,1,\ldots,N-1}$ [partially shown in
      (a)].}
   \label{fig:block_max_illustration}
\end{figure}

Since the theoretical analysis of extreme value statistics is mostly concerned
with \emph{asymptotic} distributions as discussed in
section~\ref{sect:extreme_value_statistics}, these findings are not
immediately applicable to time series $\{x_i\}_{i=1,\ldots,n}$ of finite
length $n$ which are measured in natural systems. In particular, the exact
cumulative distribution function $F(x)$ is generally unknown and, hence,
neither is $F^n(m)$ in Eq.~\eqref{eq:maximum_distribution}. Thus, one
typically has to assume that i) the given time series is a single realization
of an underlying stationary stochastic process and ii) the underlying dynamics
is ergodic such that one can use time averages instead of or in addition to
ensemble averages. This allows one to estimate the distribution of the (block)
maximum or extreme value $M_R=\max\{X_1,\ldots,X_R\}$ from the given time
series by considering non-overlapping blocks of size $R \ll n$. To be more specific, the
sequence of maximum values in disjunct blocks of length $R$ is given by
$\{m_j\}_{j=1,\ldots,[n/R]}$ where
$m_j=\max\{x_{(j-1)R+1},x_{(j-1)R+2},\ldots,x_{jR}\}$ is the maximum in block
$j$ and $[.]$ denotes the integer division; see
Fig.~\ref{fig:block_max_illustration} for an illustration. Under the above
assumptions, the estimated distribution $P_R(m)$ of the block maxima should
converge towards a specific $G(m)$ as $R\to\infty$ if a suitable sequence
$\{a_n, b_n\}$ exists as in Eq.~\eqref{eq:Fischer_Tipett_Gnedenko}. If this
convergence is sufficiently fast, one can estimate $G(m)$ reliably even for
finite $R$ (and $n$) and a finite number of realizations $N_\text{conf}$
provided that $[n/R] \times N_\text{conf} \gg 1$.

For independent symmetric $\alpha$-stable distributed processes, it is
well-known that the asymptotic limit distribution of its extreme values is the
Fr\'echet distribution in
Eq.~\eqref{eq:Frechet}~\citep{Modelling_Extremal_Events:2003} since
$\overline{P}(x)\sim x^{-1-\alpha}$. More recently, it was proved
mathematically that all self-similar S$\alpha$S processes with $\alpha < 2$ also converge in
distribution to the Fr\'echet extreme value
distribution~\citep{samorodnitsky04}. Thus, the presence of long-range memory
in the form of persistence or anti-persistence does not have any significant
influence on the asymptotic behavior in this case of stationary processes.
However, we find that there are significant finite size effects, i.e. significant
differences in the extreme value distribution for finite $R$ between the cases
with and without long-range memory.

\begin{figure}[tb]
   \centering
      \includegraphics[width=\hsize]{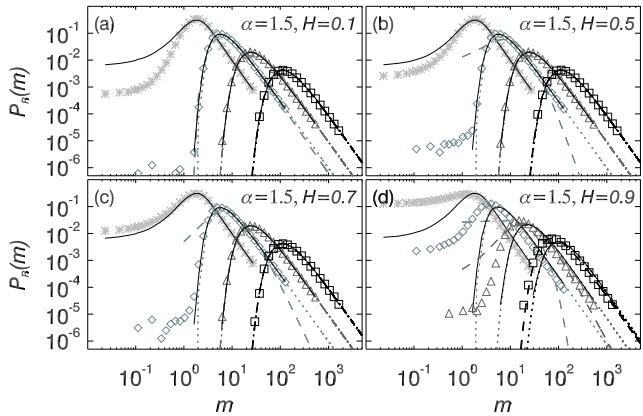}
   \caption{
   Probability density functions of block maxima for $\alpha=1.5$ and
   different values of $H$ and $R$. (a) and (b) are examples of
   anti-persistence while (c) and (d) correspond to examples of persistent
   behavior since $H=1/\alpha=2/3$ is the boundary between the two regimes.
   Shown are four different block sizes $R=10$ [stars, light gray], $R=100$
   [diamonds, gray], $R=1000$ [triangles, dark gray], and $R=10000$ [squares,
   black]. The surrogate data for the corresponding independent S$\alpha$S
   processes are indicated by solid black lines. Maximum-likelihood GEV
   fits are plotted as dotted curves (surrogate data) and dashed curves
   (original data) in the same gray tones for comparison.
   }
   \label{fig:unshuffled_vs_shuffled_bm_dist_h_panels_logaxis_1.5}
\end{figure}

\begin{figure}[tb]
   \centering
      \includegraphics[width=\hsize]{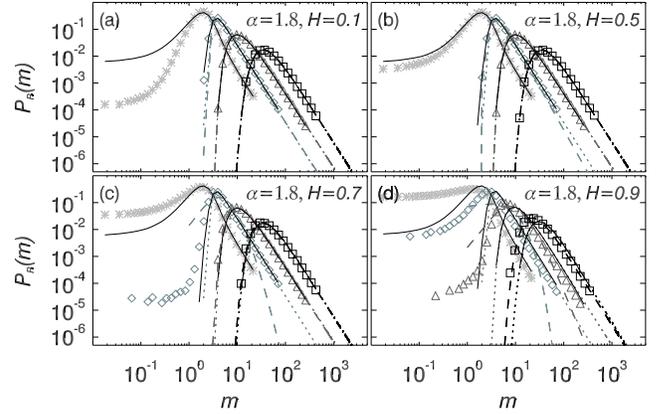}
   \caption{
   Same as Fig.~\ref{fig:unshuffled_vs_shuffled_bm_dist_h_panels_logaxis_1.5}
   but for $\alpha=1.8$. Note that now the independent case corresponds to
   $H=1/\alpha=\tfrac{5}{9}$ and we have anti-persistence for $H<\tfrac{5}{9}$
   and persistence for $H>\tfrac{5}{9}$. Thus, the persistence for $H=0.9$ is
   stronger and the anti-persistence for $H=0.1$ is weaker than in the case
   with $\alpha=1.5$ shown in
   Fig.~\ref{fig:unshuffled_vs_shuffled_bm_dist_h_panels_logaxis_1.5}.
   }
   \label{fig:unshuffled_vs_shuffled_bm_dist_h_panels_logaxis_1.8}
\end{figure}

To see this, we estimate the probability density function $P_R(m)$ of the
maxima or extreme values in non-overlapping blocks of length $R$ which are shown in
Figures~\ref{fig:unshuffled_vs_shuffled_bm_dist_h_panels_logaxis_1.5} and
\ref{fig:unshuffled_vs_shuffled_bm_dist_h_panels_logaxis_1.8} for self-similar
S$\alpha$S processes with different parameters $H$ and $\alpha$ including
persistent as well as anti-persistent cases. There are clear differences
between the original data and the independent surrogate data, both of which
were generated as described in section~\ref{sect:numerics}. These differences
are particularly pronounced for small values of $m$ and $R$ and for strongly persistent
cases ($H=0.9$). For small $m$, we observe that persistence and
anti-persistence lead to different behaviors: the probability of finding a
smaller $m$ is enhanced in the case of persistence when compared to the
surrogate data, while in the case of anti-persistence the behavior depends
sensitively on $R$. Nevertheless, in the latter case there seems to be a clear
and rapid convergence to the distribution of the independent surrogate data
with increasing $R$. Specifically, there are no significant differences for
$R>1000$.  This observation can be readily explained. Anti-persistence implies
that in an associated time series large values tend to be followed by small
values and small values are rather followed by large values which is similar
to an alternating behavior. This alternating behavior does not significantly
constrain the maximum value found in a given block if its size is sufficiently
large.  Thus, long-range memory in the form of anti-persistence does not play
an important role for the extreme value distribution even for finite $R$.

\begin{figure}[tb]
   \centering
   \includegraphics[width=\hsize]{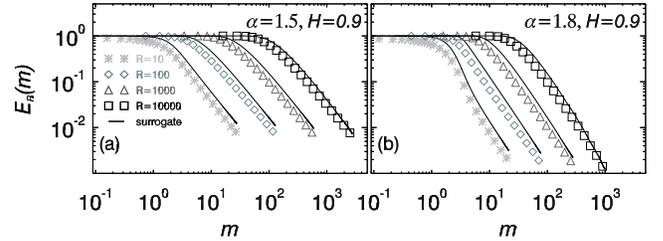}
   \caption{Exceedance probability, i.~e., the probability to find an extreme
   event larger than $m$. Shown are results for self-similar S$\alpha$S time
   series with $H=0.9$ and the corresponding independent surrogates (black
   solid lines) for different block sizes $R$ (symbols) and for different
   characteristic exponents (a) $\alpha=1.5$ and (b) $\alpha=1.8$.}
   \label{fig:unshuffled_vs_shuffled_bm_exceedance_strong_persistence}
\end{figure}

The situation is very different for the case with long-range persistence.
Since persistence corresponds to a clustering of large and small values,
respectively, more blocks will contain only smaller values such that the block
maximum will also be smaller than what would be expected based on independent
values. Thus, the distribution of block maxima will indicate a higher
probability to find smaller values compared to the independent surrogate data.
This is exactly what
Figures~\ref{fig:unshuffled_vs_shuffled_bm_dist_h_panels_logaxis_1.5} and
\ref{fig:unshuffled_vs_shuffled_bm_dist_h_panels_logaxis_1.8} show for
$H=0.9$, independent of $R$. 
As expected, the deviations from the extreme value distribution of the
surrogate data are larger for stronger persistence as a comparison of
$\alpha=1.5$ and $\alpha=1.8$ shows.  Thus, long-range memory in the form of
persistence reduces the probability of observing very large block maxima for
finite $R$. This is quantified by
Fig.~\ref{fig:unshuffled_vs_shuffled_bm_exceedance_strong_persistence}, which
shows the exceedance probability defined as
\begin{equation}
   E_R(m) = Pr(M_R>m) = \int\limits_{m}^{\infty}P_R(m^\prime)dm^\prime.
   \label{eq:cumulative_blockmax_dist}
\end{equation}
This phenomenon is not only true for self-similar S$\alpha$S processes
considered here but also for stationary long-range correlated Gaussian and
exponentially distributed processes~\citep{eichner06}.

In order to test how closely the probability density function $P_R(m)$ of the
block maxima for finite $R$ resembles its asymptotic limit given by the
generalized extreme value distribution in Eq.~\eqref{eq:gev}, we employ a
maximum likelihood estimation method -- already implemented in Matlab's
Statistics Toolbox -- to estimate its parameters and to quantify the
goodness-of-fit
\citep{MaximumLikelihoodReview_Aldrich:1997,Matlab_StatisticsToolbox}. 
The values of the shape parameters $\xi$ determining the distribution class
are reported for all considered values of $H$ and $\alpha$ in
Table~\ref{tab:shape} in the Appendix. Fig.~\ref{fig:shape_of_R} illustrates
the convergence of $\xi_R$ to $1/\alpha$ as $R\to\infty$ for selected values.
The fits themselves are shown in
Figures~\ref{fig:unshuffled_vs_shuffled_bm_dist_h_panels_logaxis_1.5} and
\ref{fig:unshuffled_vs_shuffled_bm_dist_h_panels_logaxis_1.8}. While there are
significant deviations from the GEV distribution for small $R$ and strong
persistence, we find relative agreement for $R \ge 10000$ in all cases
considered. However, we observe systematic deviations from the theoretical
value $1/\alpha$ for very large $R$. Although errors increase, the systematic
underestimation of the shape parameter is possibly not exclusively caused by
finite size effects but might be due to short-comings in the GEV fitting 
algorithm as well as the data generation algorithm.

\begin{figure}[tb]
   \centering
   \includegraphics[width=0.55\hsize]{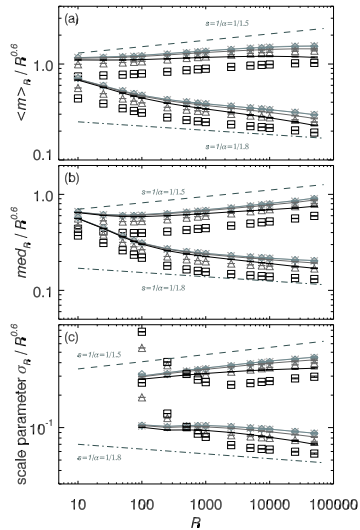}
   \caption{
   Dependence of (a) the mean $\langle m\rangle_R$ and (b) the median $med_R$
   block maximum together with (c) the scale parameter $\sigma_R$ from a GEV
   fit on the block size $R$. Shown are results for four different
   self-similar S$\alpha$S time series [gray-scale coded symbols] and corresponding
   independent surrogates [solid lines in same gray tones] --- H=0.1 [light gray,
   stars], H=0.2 [gray, diamonds], H=0.8 [dark gray, triangles], and H=0.9
   [black, squares].  Upper curves belong to $\alpha=1.5$ and lower curves to
   $\alpha=1.8$.  Shown are averages of $\langle m\rangle_R$ and $med_R$
   obtained separately for each configuration ($N_\text{conf}=1500$) together
   with the ensemble standard deviations as error bars in (a,b).  Expected
   rescaling exponents are indicated by dashed and dash-dotted lines,
   respectively. Note that the y-axis was rescaled by $R^{0.6}$ to enhance
   differences. Error bars in (c) are $95\%$-confidence intervals of
   $\sigma_R$ obtained from a GEV fit after combining all $N_\text{conf}$
   configurations to increase statistics for large $R$. 
   }
   \label{fig:mean_median_sd_r_dependence}
\end{figure}

Further evidence for the rapid convergence to the asymptotic GEV distribution
comes from the scaling of the mean, $\langle m\rangle_R$, the median, $med_R$,
and the estimated scale parameter, $\sigma_R$, with $R$ (see
Eq.~\eqref{eq:gev}), which is shown in
Fig.~\ref{fig:mean_median_sd_r_dependence}. For large $R$, the scaling
approaches the same limit for the self-similar S$\alpha$S processes and the
independent surrogate data
\footnote{Note that the differences between the different surrogate data are
partly statistical and partly due to the algorithm discussed in
section~\ref{sect:numerics}}. 
For the rescaling coefficient $a_R$ in Eq.~\eqref{eq:Fischer_Tipett_Gnedenko},
it is known that $a_R\propto R^{1/\alpha}$ asymptotically
\citep{Modelling_Extremal_Events:2003}.  This follows from studies of the
domain of attraction of the Fr\'echet class, which have proved that if the
cumulative distribution function $F(x)$ of the stochastic process decays
asymptotically as  $\overline{F}(x)\sim L(x) x^{-\alpha}$, then the rescaling
coefficient $a_R=R^{1/{\alpha}}L(R)$ for some slowly varying function
$L$~\citep{Modelling_Extremal_Events:2003}.  $L(R) \equiv constant$ for
S$\alpha$S processes.  It was also shown that centering is not necessary,
i.~e., $b_R=0$~\citep{embrechts}.

Nevertheless, there exist to our best knowledge no analytical derivations on
the convergence of $\langle m\rangle_R$, $med_R$, and $\sigma_R$. We expect a
similar behavior in the limit $R\to\infty$ where a unique Fr\'echet
distribution is approached. While we observe a scaling similar to $a_R$ for
the median in Fig.~\ref{fig:mean_median_sd_r_dependence}(b) and with
reservations for the scale $\sigma_R$ in
Fig.~\ref{fig:mean_median_sd_r_dependence}(c) there are still noticeable
deviations for the mean $\langle m \rangle_R$, and for that reason we
exclusively focus on medians in our discussion of conditional extremes, see
Sect.~\ref{sect:results_cond_block_maxima}.
Besides this, Fig.~\ref{fig:mean_median_sd_r_dependence} provides further evidence
that the asymptotic scaling of the renormalization sequence $\{a_n, b_n\}$ ---
which ensures the convergence in Eq.~\eqref{eq:Fischer_Tipett_Gnedenko} --- is
not affected by the presence of long-range memory. Yet, we also see clear
deviations from the asymptotic scaling for small $R$ as expected. In
particular, we observe a crossover in the medians at scales somewhat below
$R_\times=100$ for $\alpha=1.5$ and around $R_\times=600-700$ for $\alpha=1.8$
in the medians.  Note that this crossover is apparently independent of $H$.
This indicates that long-range memory does not play a role for the deviations
from the asymptotic scaling. Instead, the `heavier' the heavy-tailed
distribution, the smaller $R_\times$.

Based on the observed scaling behavior, it is straightforward to identify a
suitable sequence $\{a_R, b_R\}$ to ensure that
Eq.~\eqref{eq:Fischer_Tipett_Gnedenko} holds. Namely, we chose 
\begin{equation}
   a_R=R^{1/\alpha}\ \text{and}\ b_R=0\ .
   \label{eq:rescaling_properties}
\end{equation}
%
%thus, neglecting the (weak) influence of a possible correction term $L(n)$
%mentioned above. \JD{I don't think we have a non-trivial term!} 
%\AYS{Has to be at least one}

\begin{figure}[tb]
   \centering
   \includegraphics[width=\hsize]{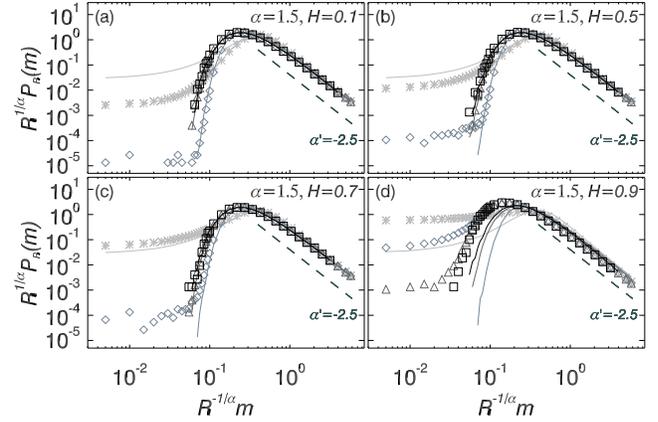}
   \caption{
   Rescaled versions of probability density functions shown in
   Fig.~\ref{fig:unshuffled_vs_shuffled_bm_dist_h_panels_logaxis_1.5}; the
   same gray tones and symbols are used. The expected asymptotic scaling of
   the tails is indicated by the dark gray dashed lines
   ($\alpha^\prime=-1-\alpha=-2.5$).
   }
   \label{fig:unshuffled_vs_shuffled_bm_dist_h_panels_collapse_logaxis_1.5}
\end{figure}

\begin{figure}[tb]
   \centering
   \includegraphics[width=\hsize]{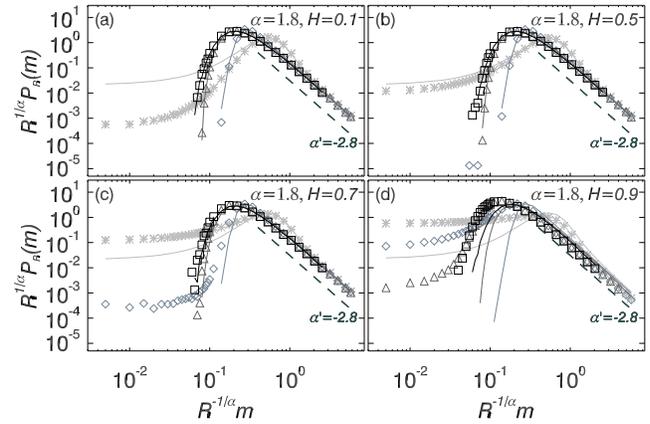}
   \caption{Same as
   Fig.~\ref{fig:unshuffled_vs_shuffled_bm_dist_h_panels_collapse_logaxis_1.5}
   but for $\alpha=1.8$. 
   }
   \label{fig:unshuffled_vs_shuffled_bm_dist_h_panels_collapse_logaxis_1.8}
\end{figure}

Using this sequence, we can rescale the probability density functions in
Figs.~\ref{fig:unshuffled_vs_shuffled_bm_dist_h_panels_logaxis_1.5} and
\ref{fig:unshuffled_vs_shuffled_bm_dist_h_panels_logaxis_1.8} according to
Eq.~\eqref{eq:Fischer_Tipett_Gnedenko} in order to obtain a scaling collapse
resembling the asymptotic distribution. This is shown in
Figs.~\ref{fig:unshuffled_vs_shuffled_bm_dist_h_panels_collapse_logaxis_1.5}
and \ref{fig:unshuffled_vs_shuffled_bm_dist_h_panels_collapse_logaxis_1.8}.
While the collapse in the tails is excellent in all considered cases, the
collapse is often not as good for small arguments. This is particularly true
for large $\alpha$ and large $H$ and confirms the findings above.

\subsection{Conditional extrema and hazard assessment}
\label{sect:results_cond_block_maxima}

As a consequence of long-range memory in self-similar S$\alpha$S processes, the
value of $X(t)$ for a given $t$ depends on the values at all earlier times.
Thus, the probability of finding a certain value $m_{j+1}$ as the block
maximum also depends on the history and, thus, the preceding block maxima
$m_{k}$ with $k \le j$. This directly allows one to make a time-dependent
hazard assessment and potentially even predict the size of future extreme
values based on the history of block maxima. In the absence of memory as for
the iid case, only \emph{time-independent} hazard assessment is possible.

\begin{figure}[tb]
   \centering
      \includegraphics[width=\hsize]{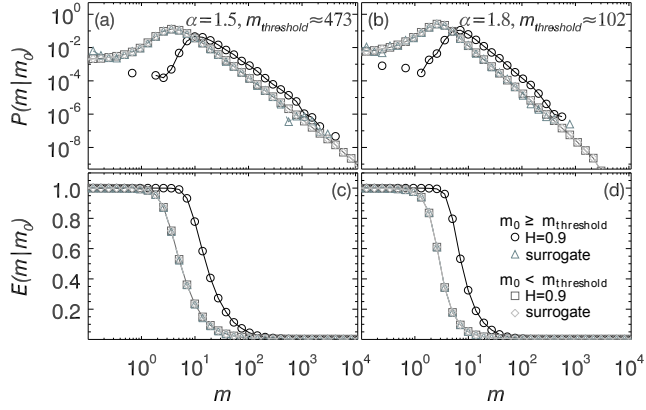}
   \caption{
   Examples of the conditional probability density functions for block maxima
   (a,b) and corresponding exceedances (c,d) for $R=100$. Left panels (a,c)
   refer to $\alpha=1.5$ and right panels (b,d) to $\alpha=1.8$.  Shown are
   results for block maxima $m$ that are preceded by a block maximum $m_0\ge
   m_\text{threshold}$ for a (large) self-similarity index $H=0.9$ [black
   circles] and for the corresponding randomized sequence of block maxima
   [gray triangles].  For comparison results for the complementary
   condition $m_0< m_\text{threshold}$ are also displayed (self-similar
   S$\alpha$S processes [dark gray squares], independent surrogate [light gray
   diamonds]). The threshold has been chosen to ensure $15000$ conditional
   maxima [(a,c) $m_\text{threshold}\approx 473$, (b,d)
   $m_\text{threshold}\approx 102$].
   }
   \label{fig:unshuffled_vs_shuffled_cond_bm_dist_compare}
\end{figure}

As one possible approach to time-dependent hazard assessment, we study the
statistics of all block maxima $m_{j+1}$ in the sequence of maxima
$\{m_j\}_{j=1,\ldots,[N/R]}$ that follow a block maximum $m_{j}=m_0$ within a
certain range --- see~\citep{schweigler11} for a related but different approach.
For example, Figs.~\ref{fig:unshuffled_vs_shuffled_cond_bm_dist_compare}(a,b)
display the conditional probability density functions of block maxima $m$ that
follow a block maximum $m_0$ larger or equal than a threshold. The thresholds
were chosen to obtain 15000 conditional maxima when combining all realizations
\footnote{This corresponds to the upper $0.999$-quantile, or 1 per mille of
all maxima.}. As can be seen, long-range persistence shifts the distributions to
the right compared to the independent surrogate case. The latter is equivalent
to the iid case. Note that the results for the inverse condition ($m_0$ is
smaller than the threshold) are also indistinguishable from the iid case.
Panels (c,d) report the corresponding exceedance probabilities, clearly
indicating clustering of extremes in the presence of long-range persistence
compared with the time-independent case, i.~e., larger events are more likely
followed by large events in the time-dependent case.

\begin{figure}[tb]
   \centering
   \includegraphics[width=\hsize]{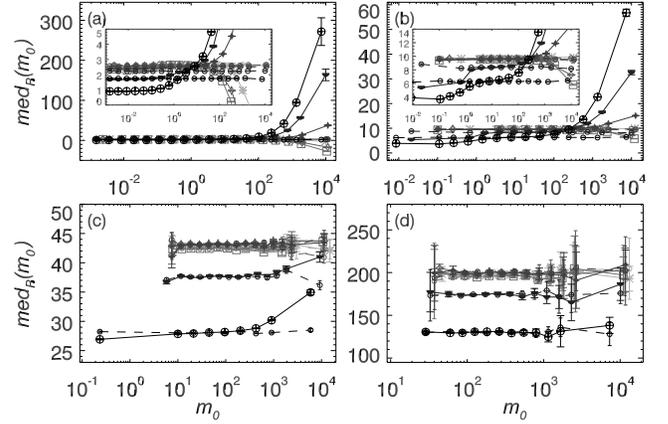}
   \caption{
   Median of the conditional distribution of block maxima,
   $\operatorname{med}_R(m_0)$, given a preceding block maximum of $m_0$ for
   $\alpha=1.5$. Results for different block sizes (a) $R=10$, (b) $R=100$,
   (c) $R=1000$, (d) $R=10000$ are shown. Different self-similarity exponents
   are both gray tone- and symbol coded; $H=0.1$ [very light gray, stars], $H=0.5$
   [light gray, squares], $H=0.6$ [gray, open diamonds], $H=0.7$ [dark gray,
   filled diamonds], $H=0.8$ [very dark gray, filled lower half circles],
   $H=0.9$ [black, circles with plus]. The values expected in the absence of
   memory effects between block maxima are shown in the same gray-scale coding but
   with smaller open circles (see text for details). Note that both
   corresponding curves for data with and without memory must intersect. Error
   bars are obtained from bootstrapping~\citep{Efron:1979,Bootstrap:1993}
   within each bin: We draw the same number of elements as in the bin with a
   cap at 50000 and consider 10000 realizations per bin. The insets in (a,b)
   show magnifications of the intersection area.
   }
   \label{fig:cond_median_h_dependence_1.5}
\end{figure}

\begin{figure}[tb]
   \centering
   \includegraphics[width=\hsize]{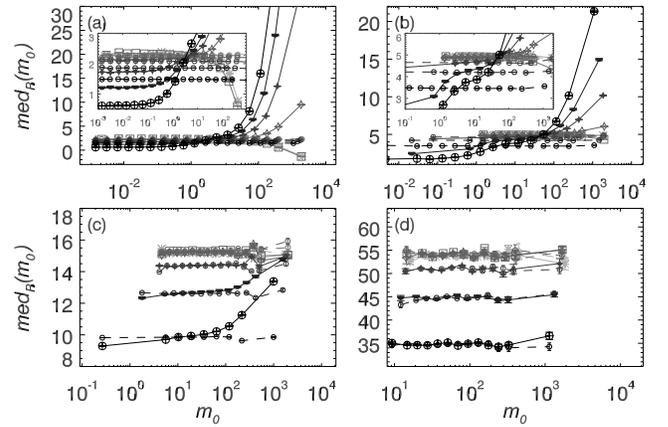}
   \caption{
   Same as Fig.~\ref{fig:cond_median_h_dependence_1.5} but for a
   characteristic exponent $\alpha=1.8$. The main panels have the same
   $x$-axes as in Fig.~\ref{fig:cond_median_h_dependence_1.5} to allow a
   better comparison.
   }
   \label{fig:cond_median_h_dependence_1.8}
\end{figure}

For a more complete understanding we now define the conditions $m_0$ as the
geometric averages of non-overlapping and exponentially growing ranges and
combine bins at both ends to ensure at least 1000 conditional maxima per
bin\footnote{We start with a set of $N_\text{bin}=20$ non-overlapping
logarithmic bins, $\{\mathfrak{M}_{k}\}_{k=1,\ldots,{N_\text{bin}}} =
\big\{[m_{k,\text{min}},m_{k,\text{max}})\big\}_k$ with $m_{k,\text{min}}\le
m_{k,\text{max}}$ and $m_{1,\text{min}}>0$, and associate the geometric mean
$m_{k,0} = (m_{k,\text{min}}m_{k,\text{max}})^{1/2}$ with the condition
$m_{k,0}$.  Due to the heavy tails in the Fr\'echet distribution, outer bins
are much less populated than bins in the center.  We therefore combine bins on
both edges, starting from $k=1$ and $k=N_\text{bin}$ and proceeding towards
the center, until at least $1000$ maxima $m_j$ are contributing to each bin.}.
Figures~\ref{fig:cond_median_h_dependence_1.5} and
\ref{fig:cond_median_h_dependence_1.8} report the full results for conditional
medians $med_R(m_{k,0})=\text{median}\{m_{j,j=2,\ldots,n} | m_{j-1}=m_{k,0}\}$
where the conditions $m_{k,0}$ are understood as ranges defined by the $k$-th
bin.
In addition to the conditional median of self-similar S$\alpha$S distributed
time series we also show the conditional medians for the randomized sequence
of block maxima for comparison. Shuffling the raw data affects the
distribution of block maxima depending on the block size $R$. Hence, for
generating independent surrogates it is here important to shuffle the sequence
of block maxima and \emph{not} the underlying time series $\{x_i\}_i$ to
preserve the distribution of block maxima. 
%In the latter case, the distribution of the maxima would
%change in most cases (small $R$; strong persistence) as follows from
%Figs.~\ref{fig:unshuffled_vs_shuffled_bm_dist_h_panels_logaxis_1.5} and
%\ref{fig:unshuffled_vs_shuffled_bm_dist_h_panels_logaxis_1.8}. 
Note that shuffling
the block maxima is equivalent to destroying long-range memory in
$\{x_i\}$ on scales larger than the block size but preserving memory on
smaller scales \citep{Schumann_MF-CMA:2011}.
%
%\footnote{It is important to shuffle the sequence of block maxima and
%\emph{not} the underlying time series $\{x_i\}_i$ to preserve the distribution
%of block maxima. In the latter case, the distribution of the maxima would
%change in most cases (small $R$; strong persistence) as follows from
%Figs.~\ref{fig:unshuffled_vs_shuffled_bm_dist_h_panels_logaxis_1.5} and
%\ref{fig:unshuffled_vs_shuffled_bm_dist_h_panels_logaxis_1.8}.}
%
Considering the surrogates corresponds to a time-independent hazard
assessment. We observe that $med_R(m_0)$ increases monotonically
with $m_0$ in the case of strong persistence, again a clear indication of the
clustering of block maxima. This dependence becomes less pronounced for larger
$R$ and indeed it should vanish in the limit $R \to \infty$. This is also true
for other processes with long-range persistence~\citep{eichner06}. In the case
of anti-persistence, the deviations from the expected behavior for independent
maxima are much smaller and only significant for small $R$. As expected, the
behavior is opposite to the case of persistence: $med_R(m_0)$ decreases
monotonically with $m_0$.

\subsection{Record Statistics}
\label{sect:results_records}

\begin{figure}[tb]
   \centering
   \includegraphics[width=\hsize]{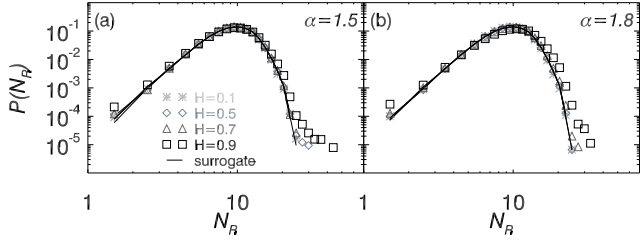}
   \caption{
   Probability of finding a given total number of records in sequences of block
   size $R=10000$ for self-similar S$\alpha$S processes with different
   parameters $\alpha$ and $H$ (symbols). Results for corresponding surrogate
   data are shown for comparison (lines).
   }
   \label{fig:record_number_of_records_distributions}
\end{figure}

\begin{figure}[tb]
   \centering
   \includegraphics[width=\hsize]{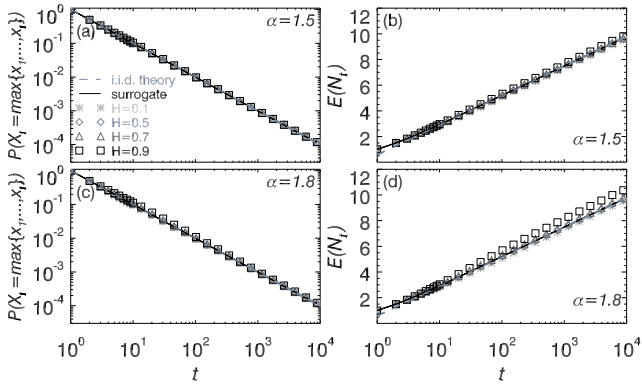}
   \caption{(a,c) Probability to break a record at a time $t$ for different
   values of $H$ and $\alpha$. Black solid lines correspond to independent
   surrogate data ($H=1/\alpha$) which don't deviate significantly from the
   theoretically expected $P(X_t=\max\{x_1,\ldots,x_t\})= 1/t$ --- see
   ~Eq.~\eqref{eq:record_rate}. (b,d) Expected total number of records
   $E(N_t)$ that occur up to time $t$ in the same color- and symbol coding as
   in (a,c). The theoretical iid behavior is $E(N_t)=ln|t|+\gamma$; see
   Eq.~\eqref{eq:expected_number_records}. Significant deviations from the iid
   behavior are visible in the presence of strong persistence ($H=0.9$,
   $\alpha=1.8$).
   }
   \label{fig:record_proabilities}
\end{figure}

While the theoretical results for record statistics presented in
section~\ref{sect:record_statistics} are all based on the assumption of iid
processes, not much is known about the case of stationary processes with
long-range memory~\citep{RecordsCorrelatedGaussianTemperatures_Newman:2010}.
For self-similar S$\alpha$S processes, we first analyze the probability
$P(N_R)$ to find $N_R$ records given a sequence of length $R$. As
Fig.~\ref{fig:record_number_of_records_distributions} shows for $R=10000$,
this probability only deviates significantly from the iid prediction --- which
is independent of the underlying distribution as follows from the discussion
in section~\ref{sect:record_statistics} --- in the case of strong persistence
($H>1/\alpha$). Since persistence implies that large values tend to follow
large values and small values rather follow small values, one expects that
there is higher probability to encounter a small number of records as well a
larger number of records compared to the iid prediction. This is exactly what
Fig.~\ref{fig:record_number_of_records_distributions} shows.
While the above results could be specific to the chosen $R$, this is not
confirmed by the dependence of the mean number of records $E(N_t)$ on $t$ with
$t \le R$. Indeed, Fig.~\ref{fig:record_proabilities} indicates that
strong deviations from the iid behavior described by
Eq.~\eqref{eq:expected_number_records} only occur in the presence of strong
persistence. While the influence of persistence on record statistics is easily
detectable in $E(N_t)$ and $P(N_R)$, the effect is more subtle if one
considers the probability to break a record at a time $t$,
$P(X_t=\max\{x_1,\ldots,x_t\})$, see Fig.~\ref{fig:record_proabilities}. This
is related to the fact, that the former are cumulative measures of the later
as follows directly from the discussion in
section~\ref{sect:record_statistics}.

Our results are consistent with earlier numerical findings for stationary
Gaussian processes with persistent long-range
memory~\citep{RecordsCorrelatedGaussianTemperatures_Newman:2010}. The authors
found that the expected number of records $E(N_t)$ for fixed $t$ increased
monotonically with increasing persistence.

\section{Application: Solar power input into the Earth's magnetosphere}
\label{sect:solar_application}

The solar wind is a prime example of plasma turbulence at low frequency
magnetohydrodynamic scales as evident from its power-law energy
distribution~\citep{zhou04}, the magnetic field
correlations~\citep{matthaeus05} and its intermittent
dynamics~\citep{burlaga01}. While understanding this type of turbulence is an
important challenge by itself, understanding the interplay between the solar
wind and the Earth's magnetosphere is another open problem of considerable
interest~\citep{Baker:2000}. The difficulty of understanding the
magnetospheric response to the solar wind variations is intimately related to
the observed range of mechanisms of energy release and multiscale coupling
phenomena~\citep{angelopoulos99,lui00,uritsky02,borovsky03,damicis07,uritsky98,chang99,chapman99,klimas00,sitnov01,consolini02,aschwanden}.
Here, we focus on the Akasofu $\epsilon$ parameter which is a solar-wind proxy
for the energy input into the Earth's magnetosphere
(see~\citep{KoskinenTanskanen:2002} for a more recent discussion). In SI units
it is defined as
\begin{equation}
  \epsilon = v \frac{B^2}{\mu_0} \ell_0^2 \sin^4(\theta/2),
  \label{eq:akasofu}
\end{equation}
where $v$ is the solar wind velocity, $B$ is the magnetic field, $\mu_0 = 4\pi
\times 10^{-7}$ is the permeability of free space, $\ell_0 \approx 7 R_E$, and
$\theta = \arctan(|B_y|/B_z)$. Geocentric solar magnetospheric (GSM)
coordinates are used.

\begin{figure}[tb]
   \centering
   \includegraphics[width=\hsize]{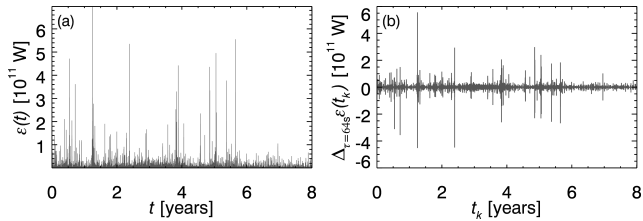}
   \caption{(a) Akasofu's $\epsilon(t)$ time series derived from ACE data
   according to Eq.~\eqref{eq:akasofu} for the years 2000 to 2007 with a
   sampling of 64s. (b)
   $\Delta_{\tau=64s}\epsilon(t_k)$ time series obeying a symmetric heavy
   tailed distribution as shown by ~\cite{CorrelatedLevy_SolarWinds_Moloney2010}.}
   \label{fig:akasofu_time_series}
\end{figure}

The ACE spacecraft~\citep{StoneETAL:1998} orbits the Earth-Sun L1 libration
point approximately $1.5 \times 10^9$m from the Earth and monitors solar wind,
interplanetary magnetic fields and high energy particles. The data can be
downloaded from {\tt http://cdaweb.gsfc.nasa.gov}. Specifically, for the years
2000-2007, we extracted the magnitude of the $x$-component of the solar wind,
and the $y$ and $z$ components of the magnetics fields, as seen respectively
by the SWEPAM and MAG instruments (level 2 72 data) of the ACE spacecraft, all
in GSM coordinates. The choice of components reflect the Poynting flux
interpretation of the $\epsilon$ parameter. For the most part, measurements
are available every $64$ and $16$s for the wind velocity and magnetic fields,
respectively. We calculated the $\epsilon$ parameter given by
Eq.~\eqref{eq:akasofu} every $64$s. Since the wind velocity and magnetic field
measurements are not synchronized, we linearly interpolated the magnetic field
measurements towards the time of the nearest wind velocity measurement. For
these $8$ years the $\epsilon$ time series consisted of $3\,944\,700$ points;
see Fig.~\ref{fig:akasofu_time_series}(a).
Measurements for wind velocities or magnetic fields are sometimes
unavailable. Approximately $9$ per cent of the points comprising the
$\epsilon$ series are missing. As in~\citep{moloney11},
we set missing data points to the value of the last valid recording
proceeding them (irrespective of the size of the data gap), 
thereby creating plateaus of constant intensity. This
minimises artifacts associated with points missing at regular
experiment-specific frequencies. We have checked that nothing changes
crucially in our statistical analyses by adopting other schemes.

\cite{watkins05} have suggested to model the $\epsilon$ parameter by a
fractional L\'evy motion as defined by
Eq.~(\ref{eq:stoev_stochastic_integral}), which implies that the series of
changes in $\epsilon$, $\Delta_\tau\epsilon(t_k)=\epsilon(t_k+\tau)-\epsilon(t_k)$
with $t_k= t_0 + k \tau$ for $k \geq 0$,
should form a self-similar S$\alpha$S process. For the $\epsilon$ time
series studied here and shown in Fig.~\ref{fig:akasofu_time_series}(b) for $\tau = 64$s, \cite{CorrelatedLevy_SolarWinds_Moloney2010} have found
that the properties of self-similarity and $\alpha$-stability both
roughly hold for 64s $\le \tau \le 4$h. Specifically, they found
$\alpha=1.55$ and $H=0.40$ indicating the presence of weak anti-persistence.
Similar values have been observed for other $\epsilon$ series 
(see~\citep{CorrelatedLevy_SolarWinds_Moloney2010} and references therein)
always indicting the presence of anti-persistence.
Given these properties, one might ask to which extent the numerical results
for time-dependent hazard assessment and records statistics of self-similar
S$\alpha$S processes presented in Section~\ref{sect:results} apply to the
$\Delta_\tau \epsilon$ series. This will be addressed in the following.

\subsection{Conditional extrema and hazard assessment}
\label{sect:eps_cond_block_maxima}

\begin{figure}[tb]
   \centering
   \includegraphics[width=\hsize]{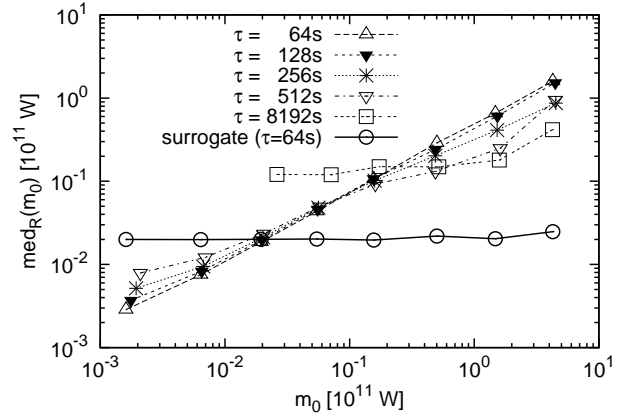}
   \caption{Median of the conditional distribution of block maxima
   $\operatorname{med}_R(m_0)$ given a preceding block maximum of
   $m_0$ within a logarithmic bin for the $\Delta_\tau \epsilon$
   series with different values of 
   $\tau$ [symbols and line style coded]. The block size is $R=100$.
   Surrogate data (obtained by shuffling the $\Delta_\tau \epsilon=64s$
   series' block maxima) corresponding to the iid case are shown for comparison [solid
   curve, circles]. Surrogates for other values of $\tau$ (not shown) are
   similar but increase monotonically with $\tau$.}
   \label{fig:eps_median}
\end{figure}

Fig.~\ref{fig:eps_median} shows that the median of the conditional
distribution of the block maxima, $\operatorname{med}_R(m_0)$, increases
monotonically and approximately as a power law with $m_0$ for $\tau=64$s.  As
$\tau$ is increased the power law's exponent decreases, eventually leaving the
power law regime, and approaching the iid case (median independent of
condition) in the limit of very large $\tau$. Note that the median values of
the block maxima and their corresponding surrogates, obtained by shuffling the
block maxima, naturally increase for
larger block sizes since the fewer blocks are more and more dominated by the
largest extreme values in the time series (cp. $\tau=8192$s). Our findings
indicate clustering of extreme values for not too large $\tau$ and, thus, persistence --- in sharp
contrast to what is expected for anti-persistent self-similar S$\alpha$S
processes (see
Figs.~\ref{fig:cond_median_h_dependence_1.5},~\ref{fig:cond_median_h_dependence_1.8}).
This contradiction provides clear evidence that only some properties of the
$\Delta_\tau \epsilon$ series can be (roughly) characterized by a self-similar
S$\alpha$S process.  In particular, the memory reflected in the $\Delta_\tau
\epsilon$ is not reliably captured by a self-similar S$\alpha$S process ---
potentially due to non-stationarities. A similar conclusion has been reached
by ~\cite{CorrelatedLevy_SolarWinds_Moloney2010} who investigated the
distributions of block maxima of $\Delta_\tau \epsilon$. Nevertheless, the
findings summarized by Fig.~\ref{fig:eps_median} show unambiguously that
time-dependent hazard assessment can be successfully applied to the considered
time series and clearly outperforms any time-independent hazard assessment of
future extreme values.

\subsection{Record Statistics}
\label{sect:eps_records}

\begin{figure}[tb]
   \centering
   \includegraphics[width=\hsize]{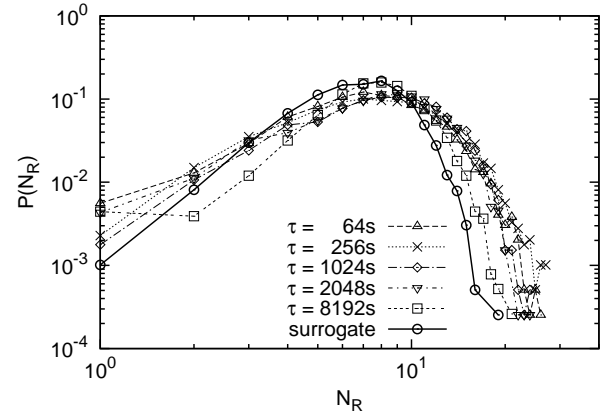}
   \caption{Probability density function for the number of records in
   non-overlapping blocks
   of size $R=1000$ of the $\Delta_\tau \epsilon$ series for different values
   of $\tau$ and for surrogate data (obtained by shuffling the $\Delta_\tau \epsilon$
   series) corresponding to the iid case are shown for comparison.}
   \label{fig:eps_rdist}
\end{figure}

\begin{figure}[tb]
   \centering
   \includegraphics[width=\hsize]{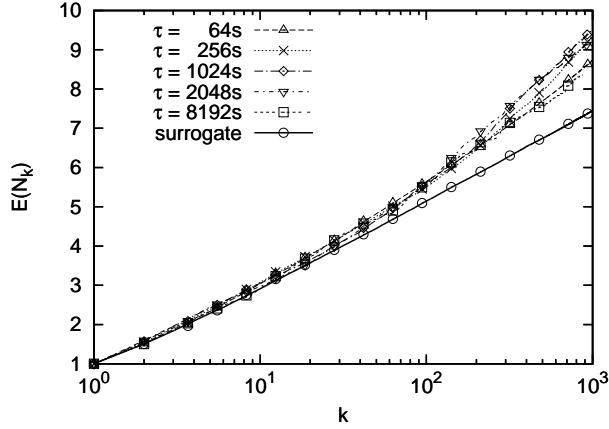}
   \caption{Expected number of records as a function of values $k$ for
   the $\Delta_\tau \epsilon$ series for different values of 
   $\tau$. Estimates
   are based on non-overlapping blocks of size $R=1000$. Surrogate data
   (obtained by shuffling the $\Delta_\tau \epsilon$ series) corresponding to
   the iid case are shown for comparison.}
   \label{fig:eps_rmean}
\end{figure}

\begin{figure}[tb]
   \centering
   \includegraphics[width=\hsize]{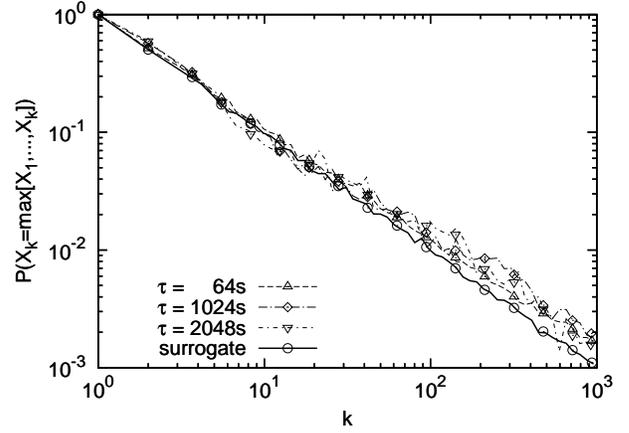}
   \caption{Probability to observe a record at values $k$ for the
   $\Delta_\tau \epsilon$ series for different values of 
   $\tau$. Estimates are
   based on non-overlapping blocks of size $R=1000$. Surrogate data (obtained
   by shuffling the $\Delta_\tau \epsilon$ series) corresponding to the iid
   case are shown for comparison.}
   \label{fig:eps_tprob}
\end{figure}

The observed clustering in the $\Delta_\tau \epsilon$ series discussed in
section~\ref{sect:eps_cond_block_maxima} is also present in the record
statistics. Fig.~\ref{fig:eps_rdist} shows that the probability $P(N_R)$ to
find $N_R$ records in a sequence of length $R$ is much higher for large $N_R$
than what is expected for the iid case. This is further confirmed by
Fig.~\ref{fig:eps_rmean} which shows that the expected number of records grows
much faster than in the iid case.  Both figures also suggest a dependence on
$\tau$. In particular, the deviations from the iid case for large times
increase monotonically with $\tau$ until they reach a maximum around $\tau
=1024$s. As $\tau$ is increased further the deviations decrease monotonically
with $\tau$ and finally they approach the iid case in the limit of very large
$\tau$.
This is supported by the probability to observe a record at an elapsed time
$t$ shown in Fig.~\ref{fig:eps_tprob}. Note that in contrast to these results
found for records the conditional maxima in Fig.~\ref{fig:eps_median}
apparently showed a monotonic dependence on $\tau$.

\subsection{Discussion}
\label{sect:eps_discussion}

\begin{figure}[tb]
   \centering
   \includegraphics[width=\hsize]{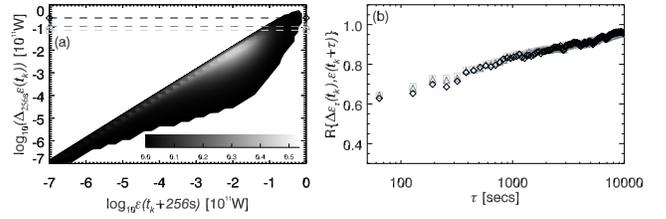}
   \caption{(a) Probability density function to observe a given pair of
values $\epsilon(t_k+\tau)$ and $\Delta_\tau\epsilon(t_k)=\epsilon(t_k+\tau)-\epsilon(t_k)$ 
simultaneously for a fixed value of $\tau=256$s and only for pairs 
where $\Delta_\tau\epsilon(t_k)>0$; logarithmic
bins were used. The dashed lines indicate from top to bottom the 99.9\% [black,
diamonds], 99.5\% [gray, triangles], and the 99\% [light gray, squares] quantile of
the $\Delta_{256\text{s}}\epsilon$ distribution. (b) Pearson's correlation
coefficient calculated for the same upper quantiles as in (a) [same symbols
and gray tones] for all time series pairs
$\big(\Delta_\tau\epsilon(t_k),\epsilon(t_k+\tau)\big)$ as a function of the time
shift $\tau$.
   }
   \label{fig:akasofu_correlation}
\end{figure}

While we have focused here on $\Delta_\tau \epsilon$ series, our results
with respect to time-dependent hazard assessment are directly applicable 
to the $\epsilon$ series itself. As Fig.~\ref{fig:akasofu_correlation} shows,
there is a strong correlation between $\Delta_\tau\epsilon(t_k)$ and 
$\epsilon(t_k+\tau)$ --- as measured by the Pearson correlation coefficient
--- if one considers the largest values in $\Delta_\tau\epsilon(t_k)$. This
implies that a large extreme value in $\Delta_\tau\epsilon$ typically 
indicates a large value in $\epsilon$, while a smaller extreme value 
in $\Delta_\tau\epsilon$ tends to correspond to a smaller value in $\epsilon$
--- independent of the exact value of $\tau$.
Thus, one can translate our findings from Section~\ref{sect:eps_cond_block_maxima}:
Extreme values in $\epsilon$ tend to be clustered, which makes time-dependent hazard
assessment in the context of space weather feasible. These results are just a first 
step and clearly need to be investigated in more detail in future work.

\section{Summary}
\label{sect:conclusion}

In summary, our numerical analysis of self-similar symmetric $\alpha$-stable
processes has shown that the presence of long-range memory can lead to
significant finite size effects in extreme value and record statistics despite
the fact that some of the asymptotic behavior is not affected. The finite size
effects are particularly pronounced if the long-range memory leads to
persistent behavior. We also found that long-range memory allows a
time-dependent hazard assessment of the size of future extreme events based on
extreme events observed in the past. Time-dependent hazard assessment based on
the value of the previous block maximum significantly outperforms
time-independent hazard assessment and, thus, provides a significant
improvement. 

Moreover, we showed that such a time-dependent hazard assessment is directly
applicable in the context of the solar power influx into the magnetosphere.
Since many processes in nature are characterized by long-range memory and
heavy-tailed distributions --- especially in space physics
\citep{watkins05,CorrelatedLevy_SolarWinds_Moloney2010} --- our findings
are of general interest. Further examples outside geophysics include communications
traffic~\citep{laskin02}.

%%% End of body of article:

%%%%%%%%%%%%%%%%%%%%%%%%%%%%%%%%
%% Optional Appendix goes here
%
%%%%%%%%%%%%%%%%%
% Geophysical Research Letters only allows an appendix without a letter.
%% You can get this result with
%  \section*{Appendix}
%  or
%  \section*{Appendix: Title}
%%%%%%%%%%%%%%%%%
%
% \appendix resets counters and redefines section heads
% but doesn't print anything.
% After typing  \appendix
%
% \section{Here Is Appendix Title}
% will print
% Appendix A: Here Is Appendix Title
%
% \section*{Appendix}
% will print
% Appendix
%
% \section*{Appendix: Here Is Appendix Title}
% will print
% Appendix: Here Is Appendix Title
%
% For only 1 appendix \appendix \section{Appendix} is preferred.
% which will print
% Appendix A

\appendix
\section*{Appendix}

\begin{table}[p]
   % Has to be removed by publisher!!!
   \scriptsize
   % Has to be removed by publisher!!!
\caption{Shape parameters $\xi$ and corresponding maximum error obtained from
$95\%$-confidence intervals of a maximum likelihood fit of
Eq.~\eqref{eq:gev} to the data. Note that for $R\le 100$ there was often no
convergence in the estimation meaning the probability density cannot be
properly approximated by the GEV distribution in Eq.~\eqref{eq:gev}. Positive values of $\xi$ indicate a Fr\'echet distribution. Asymptotically, $\xi$ should approach $1/\alpha$ based on theoretical results~\citep{samorodnitsky04}.}
%\begin{tabular*}{0.8\hsize}{@{\extracolsep{\fill}}|l||c|c||c|c|}
\begin{tabular}{|l||c|c||c|c|}
   \hline
   & \multicolumn{2}{c||}{$\alpha=1.5$} & \multicolumn{2}{c|}{$\alpha=1.8$}\\
   \hline
   & self-similar S$\alpha$S & shuffled S$\alpha$S & self-similar S$\alpha$S & shuffled S$\alpha$S\\
   & shape $\xi$             & shape $\xi$         & shape $\xi$             & shape $\xi$        \\
   \hline
   H=0.1   &                     &                     &                     &                    \\
   R=1000  & $0.6679 \pm 0.0017$ & $0.6657 \pm 0.0017$ & $0.5751 \pm 0.0017$ & $0.5734 \pm 0.0017$\\
   R=10000 & $0.6582 \pm 0.0054$ & $0.6532 \pm 0.0054$ & $0.5489 \pm 0.0051$ & $0.5448 \pm 0.0050$\\
   \hline
   H=0.2   &                     &                     &                     &                    \\
   R=1000  & $0.6682 \pm 0.0017$ & $0.6653 \pm 0.0017$ & $0.5756 \pm 0.0017$ & $0.5738 \pm 0.0017$\\
   R=10000 & $0.6575 \pm 0.0054$ & $0.6510 \pm 0.0054$ & $0.5491 \pm 0.0051$ & $0.5443 \pm 0.0050$\\
   \hline
   H=0.3   &                     &                     &                     &                    \\
   R=1000  & $0.6682 \pm 0.0017$ & $0.6647 \pm 0.0017$ & $0.5763 \pm 0.0017$ & $0.5738 \pm 0.0017$\\
   R=10000 & $0.6573 \pm 0.0054$ & $0.6509 \pm 0.0054$ & $0.5498 \pm 0.0051$ & $0.5451 \pm 0.0051$\\
   \hline
   H=0.4   &                     &                     &                     &                    \\
   R=1000  & $0.6683 \pm 0.0017$ & $0.6649 \pm 0.0017$ & $0.5764 \pm 0.0017$ & $0.5739 \pm 0.0017$\\
   R=10000 & $0.6584 \pm 0.0054$ & $0.6501 \pm 0.0054$ & $0.5501 \pm 0.0051$ & $0.5473 \pm 0.0051$\\
   \hline
   H=0.5   &                     &                     &                     &                    \\
   R=1000  & $0.6685 \pm 0.0017$ & $0.6654 \pm 0.0017$ & $0.5761 \pm 0.0017$ & $0.5756 \pm 0.0016$\\
   R=10000 & $0.6589 \pm 0.0054$ & $0.6513 \pm 0.0054$ & $0.5515 \pm 0.0051$ & $0.5496 \pm 0.0051$\\
   \hline
   H=0.6   &                     &                     &                     &                    \\
   R=1000  & $0.6684 \pm 0.0017$ & $0.6667 \pm 0.0017$ & $0.5751 \pm 0.0017$ & $0.5756 \pm 0.0017$\\
   R=10000 & $0.6608 \pm 0.0054$ & $0.6556 \pm 0.0054$ & $0.5533 \pm 0.0051$ & $0.5488 \pm 0.0051$\\
   \hline
   H=0.7   &                     &                     &                     &                    \\
   R=1000  & $0.6681 \pm 0.0017$ & $0.6675 \pm 0.0017$ & $0.5742 \pm 0.0016$ & $0.5714 \pm 0.0017$\\
   R=10000 & $0.6625 \pm 0.0054$ & $0.6568 \pm 0.0054$ & $0.5530 \pm 0.0051$ & $0.5418 \pm 0.0051$\\
   \hline
   H=0.8   &                     &                     &                     &                    \\
   R=1000  & $0.6656 \pm 0.0017$ & $0.6586 \pm 0.0017$ & $0.5496 \pm 0.0011$ & $0.5653 \pm 0.0017$\\
   R=10000 & $0.6622 \pm 0.0054$ & $0.6431 \pm 0.0054$ & $0.5519 \pm 0.0051$ & $0.5296 \pm 0.0051$\\
   \hline
   H=0.9   &                     &                     &                     &                    \\
   R=1000  & $0.4199 \pm 0.0017$ & $0.6410 \pm 0.0006$ & $0.2716 \pm 0.0005$ & $0.5510 \pm 0.0017$\\
   R=10000 & $0.6613 \pm 0.0054$ & $0.6202 \pm 0.0054$ & $0.5484 \pm 0.0051$ & $0.5108 \pm 0.0050$\\
   \hline
\end{tabular}
\label{tab:shape}
\end{table}

%%%%%%%%%%%%%%%%%%%%%%%%%%%%%%%%%%%%%%%%%%%%%%%%%%%%%%%%%%%%%%%%
%
% Optional Glossary or Notation section, goes here
%
%%%%%%%%%%%%%%
% Glossary only allowed in Reviews of Geophysics
% \section*{Glossary}
% \paragraph{Term}
% Term Definition here
%
%%%%%%%%%%%%%%
% Notation -- End each entry with a period.
% \begin{notation}
% Term & definition.\\
% Second Term & second definition.
% \end{notation}
%%%%%%%%%%%%%%%%%%%%%%%%%%%%%%%%%%%%%%%%%%%%%%%%%%%%%%%%%%%%%%%%
%
%  ACKNOWLEDGMENTS

\begin{acknowledgments}
   We thank A. Rasmussen for helpful discussions. This project was supported by Alberta Innovates (formerly Alberta Ingenuity).
\end{acknowledgments}

%% ------------------------------------------------------------------------ %%
%
%  REFERENCE LIST AND TEXT CITATIONS
%
% Either type in your references using
% \begin{thebibliography}{}
% \bibitem{}
% Text
% \end{thebibliography}
%
% Or,
%
% If you use BiBTeX for your References, please produce your .bbl
% file and copy the contents into your paper here.
%
% Follow these steps:
% 1. Run LaTeX on your LaTeX file.
%
% 2. Run BiBTeX on your LaTeX file.
%
% 3. Open the new .bbl file containing the reference list and
%   copy all the contents into your LaTeX file here.
%
% 4. Comment out the old \bibliographystyle and \bibliography commands.
%
% 5. Run LaTeX on your new file before submitting.
%
% AGU does not want a .bib or a .bbl file, but asks that you
% copy in the contents of your .bbl file here.

%\bibliographystyle{agu}
%\bibliography{references,j2,articles,books}

% AGU later needs bbl copied here

% Fig 1

% Fig 2

% Fig 3

% Fig 4

% Fig 5

% Fig 6
\begin{figure}[tb]
   \centering
   \includegraphics[width=\hsize]{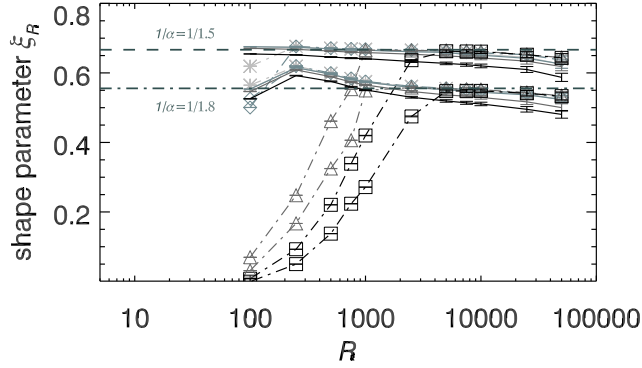}
   \caption{Convergence of the shape parameter $\xi_R$ towards the theoretical
   value $1/\alpha$ as $R\to\infty$ obtained from GEV fits for four different
   self-similar S$\alpha$S time series [gray-scale coded dash-dotted lines and
   symbols] and corresponding independent surrogates [solid lines in same
   gray] --- H=0.1 [light gray, stars], H=0.2 [gray, diamonds], H=0.8 [dark
   gray, triangles], and H=0.9 [black, squares]. Upper curves correspond to
   $\alpha=1.5$ while lower curves belong to $\alpha=1.8$. Error bars are
   $95\%$-confidence intervals of $\xi_R$ obtained from the GEV fit after
   combining the data of all 1500 configurations.
   }
   \label{fig:shape_of_R}
\end{figure}

% Fig 7

% Fig 8

% Fig 9

% Fig 10

% Fig 11

% Fig 12

% Fig 13

% Fig 14

% Fig 15 (new Figure)

% Fig 16 (old 15)

% Fig 17 (old 16)

% Fig 18 (old 17)

% Fig 19 (old 18)

% Fig 20 (new Figure)

%% ------------------------------------------------------------------------ %%
%
%  END ARTICLE
%
%% ------------------------------------------------------------------------ %%

\end{article}

%% Enter Figures and Tables here:

% When submitting articles through the GEMS system:
% COMMENT OUT ANY COMMANDS THAT INCLUDE GRAPHICS.

% Figure captions go below this illustration; Table captions go above tables

% ONE-COLUMN figure/table, including eps graphics
%
% \begin{figure}
% \noindent\includegraphics[width=20pc]{samplefigure.eps}
% \caption{Caption text here}
% \end{figure}
% \end{document}
%
% \begin{table}
% \caption{}
% \end{table}
%
% ---------------
% TWO-COLUMN figure/table
%
% \begin{figure*}
% \noindent\includegraphics[width=39pc]{samplefigure.eps}
% \caption{Caption text here}
% \end{figure*}
%
% \begin{table*}
% \caption{Caption text here}
% \end{table*}
%
% see below for how to make landscape figures or tables

%%% End the article here:

\end{document}